\documentclass[prl,twocolumn,superscriptaddress,nofootinbib]{revtex4}
\usepackage[dvipdfmx,hiresbb]{graphicx}
\usepackage{color}
\usepackage{enumerate}
\usepackage{epsfig}
\usepackage{enumitem}
\usepackage{amsmath,amssymb,latexsym}
\usepackage{ascmac}
\usepackage{bm}
\newtheorem{theorem}{{\bf Theorem}}

\newtheorem{lemma}{{\bf Lemma}}

\newcommand{\sq}{\qquad $\blacksquare$}

\def\U#1{{\rm #1}} 

\newcommand{\red}{\textcolor{red}}

\newcommand{\bra}[1]{\langle #1 |}
\newcommand{\ket}[1]{| #1 \rangle}

\newcommand{\expect}[1]{\left\langle #1 \right\rangle} 

\def\vac{\U{vac}}
\def\sample{\U{sample}}
\def\Pr{\U{Pr}}
\def\bl{\U{block}}
\def\obs{\U{obs}}
\def\odd{\U{odd}}
\def\even{\U{even}}
\def\wt{\U{wt}}
\def\ph{\U{ph}}
\def\em{\U{em}}
\def\bit{\U{bit}}
\def\th{\U{th}}
\def\tr{\U{tr}}

\def\vac{{\rm vac}}

\begin{document}
\title{
Quantum key distribution with simply characterized light sources
}
\author{Akihiro Mizutani$^{\ast}$}
\affiliation{Mitsubishi Electric Corporation, Information Technology R\&D Center, 5-1-1 Ofuna, Kamakura-shi, Kanagawa, 247-8501 Japan}
\author{Toshihiko Sasaki}
\affiliation{Photon Science Center, Graduate School of Engineering, The University of Tokyo, Bunkyo-ku, Tokyo 113-8656, Japan}
\author{Yuki Takeuchi}
\affiliation{
NTT Communication Science Laboratories, NTT Corporation, 
3-1 Morinosato Wakamiya, Atsugi-shi, Kanagawa 243-0198, Japan}
\author{Kiyoshi Tamaki}
\affiliation{
  Graduate School of Science and Engineering for Research, University of Toyama, Gofuku 3190, Toyama, 930-8555, Japan\\
${}^{\ast}${\it Mizutani.Akihiro@dy.MitsubishiElectric.co.jp}
}
\author{Masato Koashi}
\affiliation{Photon Science Center, Graduate School of Engineering, The University of Tokyo, Bunkyo-ku, Tokyo 113-8656, Japan}
\begin{abstract}
  To guarantee the security of quantum key distribution (QKD), several assumptions on light sources must be satisfied. 
  For example, each random bit information is precisely encoded on an optical pulse and the photon-number probability 
  distribution of the pulse is exactly known.
  Unfortunately, however, it is hard to check if all the assumptions are really met in practice, and it is preferable that
  we have minimal number of device assumptions.
In this paper, we adopt the differential-phase-shift (DPS) QKD protocol and drastically mitigate the requirements on light sources. 
Specifically, we only assume the independence among emitted pulses, the independence of the vacuum emission probability
from a chosen bit, and
upper bounds on the tail distribution function of the total photon number in a single block of pulses for single, two and three photons. 
Remarkably, no other detailed characterizations, such as the amount of phase modulation, are required. 
Our security proof significantly relaxes demands for light sources,
which paves a route to guarantee implementation security with simple verification of the devices.
\end{abstract}
\maketitle

{\large \bf Introduction}\\
Quantum key distribution (QKD) holds promise for information-theoretically secure communication between two distant parties, Alice and
Bob~\cite{LoNphoto2014}.
Since QKD is a physical cryptography in which security is based on a mathematical model of the devices,
several assumptions on Alice's light source and Bob's measurement unit have to be satisfied to guarantee the security. 
Any discrepancies between the device model and properties of the actual devices could be exploited to hack the implemented QKD systems.
In fact, several experiments to crack implementation of
QKD systems have been reported~\cite{PhysRevA.91.032326,PhysRevA.92.022304,Lydersen2010,Gerhardt2011},
which is a crucial threat for security of QKD, and therefore it is important to close the gap between theory and practice.

So far, tremendous efforts have been made to relax the demands for light sources
(see e.g. a review article~\cite{Diamanti16}). 
One possible approach to close the gap is to use device-independent (DI) QKD (see e.g.~\cite{Rotem18} and references therein).
However, to share an almost maximally entangled state between distant Alice and Bob 
in order to violate the Bell inequality is not practical with current technology, which renders DI protocols impractical at present. 
On the other hand, as for the device-dependent QKD, the BB84 protocol~\cite{bb84} is one of the most investigated protocols,
and its security assuming an ideal single-photon source~\cite{PhysRevLett.85.441,Tomamichel12,Tomamichel2017}
and an ideal phase randomized coherent-light source~\cite{PhysRevA.89.022307} were proved.
The security proofs were generalized to accommodate dominant imperfections of the devices.
For example, perfect phase randomization is relaxed to discrete phase randomization~\cite{1367-2630-17-5-053014},
inter-pulse intensity correlations between neighboring pulses have been accommodated~\cite{qcrypt2017tomita}, and 
a perfectly symmetric encoding of random bit information is relaxed to asymmetric encoding with
the loss-tolerant protocol~\cite{PhysRevA.90.052314,AkihiroSCIC2018}. 
Another promising protocol is the round-robin differential phase shift (RRDPS) protocol~\cite{Sasaki2014}.
Its implementation security proof has been studied~\cite{PhysRevA.92.060303}, which shows that the RRDPS protocol 
is robust against source flaws. 
However, the variable-delay interferometer used in this protocol is an obstacle to simple implementation. 

In this paper, we adopt the differential-phase-shift (DPS) protocol~\cite{dps2003} and drastically mitigate the demands for light sources, 
which is useful for simple characterization of the devices with quantified security.
Our characterizations of light sources are based on the photon-number statistics of emitted pulses.
Specifically, we suppose that the vacuum emission probability of each pulse is independent of a chosen bit, and an upper bound 
$q_n$ (for $n\in\{1,2,3\}$) on the probability that each block of pulses contains $n$ or more photons. 
Here, these probabilities are the ones that would be obtained if we performed a photon number measurement,
and we do not assume that the state is a classical mixture of Fock states.
Remarkably, 
detailed characterizations of the source devices that were needed in previous security proofs of DPS
protocol~\cite{Kiyoshi2012dps,Akihiro2017QSTDPS} and the original DPS protocol~\cite{dps2003},
such as the precise control of phase modulations, complete knowledge of the photon-number probability 
distribution, block-wise phase randomization, and a single-mode assumption on the emitted pulse are not necessary.
\\\\{\large \bf Results}\\
{\bf Assumptions  on the devices.}
Before describing the protocol, we summarize the assumptions we make on the source and the receiver. 
First, we list up the assumptions on Alice's source as follows.
In this paper, for simplicity of the security analysis, 
we consider the case where Alice employs three pulses contained in a single-block.
\begin{enumerate}[label=(A\arabic*)]
\setlength{\parskip}{0cm}
\setlength{\itemsep}{0cm}
\item
  Alice chooses a random three-bit sequence $\vec{b}_A:=b^A_1b^A_2b^A_3\in\{0,1\}^3$, and $b^A_i$ is encoded only on 
  the $i^{\th}$ pulse in system $S_i$. 
  Depending on the chosen $\vec{b}_A$, Alice prepares a following three-pulse state
  in system $S:=S_1S_2S_3$: 
  \begin{align}
    \hat{\rho}^{\vec{b}_A}_S:=\bigotimes^3_{i=1}\hat{\rho}^{b^A_i}_{S_i}.
\end{align}
  Here, $\hat{\rho}^{b^A_i}_{S_i}$ denotes a density operator of the $i^{\th}$ pulse when $b^A_i$ is chosen.
  We suppose that each system $R_i$ that purifies each of the state $\hat{\rho}^{b^A_i}_{S_i}$ is possessed by Alice,
  and Eve does not have access to system $R_i$.
    \item
      The vacuum emission probability of the $i^{\th}$ pulse is independent of the chosen bit $b^A_i$. That is,
      we require that the following equality holds for any $i$:
      \begin{align}
        \tr\hat{\rho}^{0}_{S_i}\ket{\vac}\bra{\vac}=\tr\hat{\rho}^{1}_{S_i}\ket{\vac}\bra{\vac},
        \label{vacbitindep}
      \end{align}
      where $\ket{\vac}$ is the vacuum state. 
    \item
      For any chosen bit sequence $\vec{b}_A$, 
      the probability that a single-block of pulses contains $n$ (with $n\in\{1,2,3\})$ or more photons is upper-bounded by $q_n$.
      That is,
      \begin{align}
        \Pr\{n_{\bl}\ge n\}\le q_n,
        \label{qn}
      \end{align}
      where $n_{\bl}$ denotes the number of photons contained in a single-block
            \footnote{Note that $n_{\bl}$ is the sum of the number of photons in all the optical modes. }.
            By using a calibration method based on a conventional Hanbury-Brown-Twiss setup with threshold photon detectors~\cite{Kumazawa2017},
            Alice can verify $\{q_n\}^3_{n=1}$ before running the protocol. 
            If $\{q_n\}^3_{n=1}$ are estimated from such an off-line test, we need to assume that these bounds do not change during 
            the on-line experiment.

\end{enumerate}
  We emphasize that for the security proof, we do not make any assumptions on phase modulations. That is, 
  the precise control over the phase modulation and its characterization are not needed. 
  We also emphasize that we do not make the single-mode assumption on the pulses,
  and the optical mode of the emitted pulse can depend on the bit $b^A_i$.
  This includes, for example, the case where the state of the pulse when $b^A_i=0$ (1) is 
  horizontal (vertical) polarization state.
  Our framework covers the original DPS protocol~\cite{dps2003} using coherent states $\{\ket{\alpha},\ket{-\alpha}\}$.
  In this case, $q_n$ in Eq.~(\ref{qn}) is obtained through a priori Poissonian assumption.
  We note that the previous security proofs~\cite{Kiyoshi2012dps,Akihiro2017QSTDPS} of the DPS protocol have assumed
  ideally phase modulated single-mode coherent states $\{\ket{\alpha},\ket{-\alpha}\}$ with block-wise phase randomization,
  which is removed in our analysis.
\\\\
Next, we list up the assumptions on Bob's measurement as follows.
\begin{enumerate}[label=(B\arabic*)]
\setlength{\parskip}{0cm}
\setlength{\itemsep}{0cm}
\item
Bob uses a one-bit delay Mach-Zehnder interferometer with two 
50:50 beam splitters (BSs) and with its delay being equal to the interval of the neighboring emitted pulses.
\item
  After the interferometer, the pulses are detected by two 
  photon-number-resolving (PNR) detectors, which can discriminate the vacuum, a single-photon, and two or more photons of a specific
  optical mode. A click event of each detector corresponds to bit values of 0 and 1, respectively.
  We suppose that the quantum efficiencies and dark countings are the same for both detectors. 
\end{enumerate}
In Bob's measurement, the $j^{\th}$ ($1\le j\le 2$) time slot is defined as an 
expected detection time at Bob's detectors from the superposition of the $j^{\th}$ and $(j+1)^{\th}$ incoming pulses.
Also, the $0^{\th}$ ($3^{\U{rd}}$) time slot is defined as an expected detection time
at Bob's detectors from the superposition of the 1$^{\U{st}}$ ($3^{\U{rd}}$) incoming pulse and the $3^{\U{rd}}$ incoming pulse in the previous block
(1$^{\U{st}}$ incoming pulse in the next block).
\\\\{\bf Protocol.}
\begin{figure}[t]
  \centering
\includegraphics[width=8.5cm]{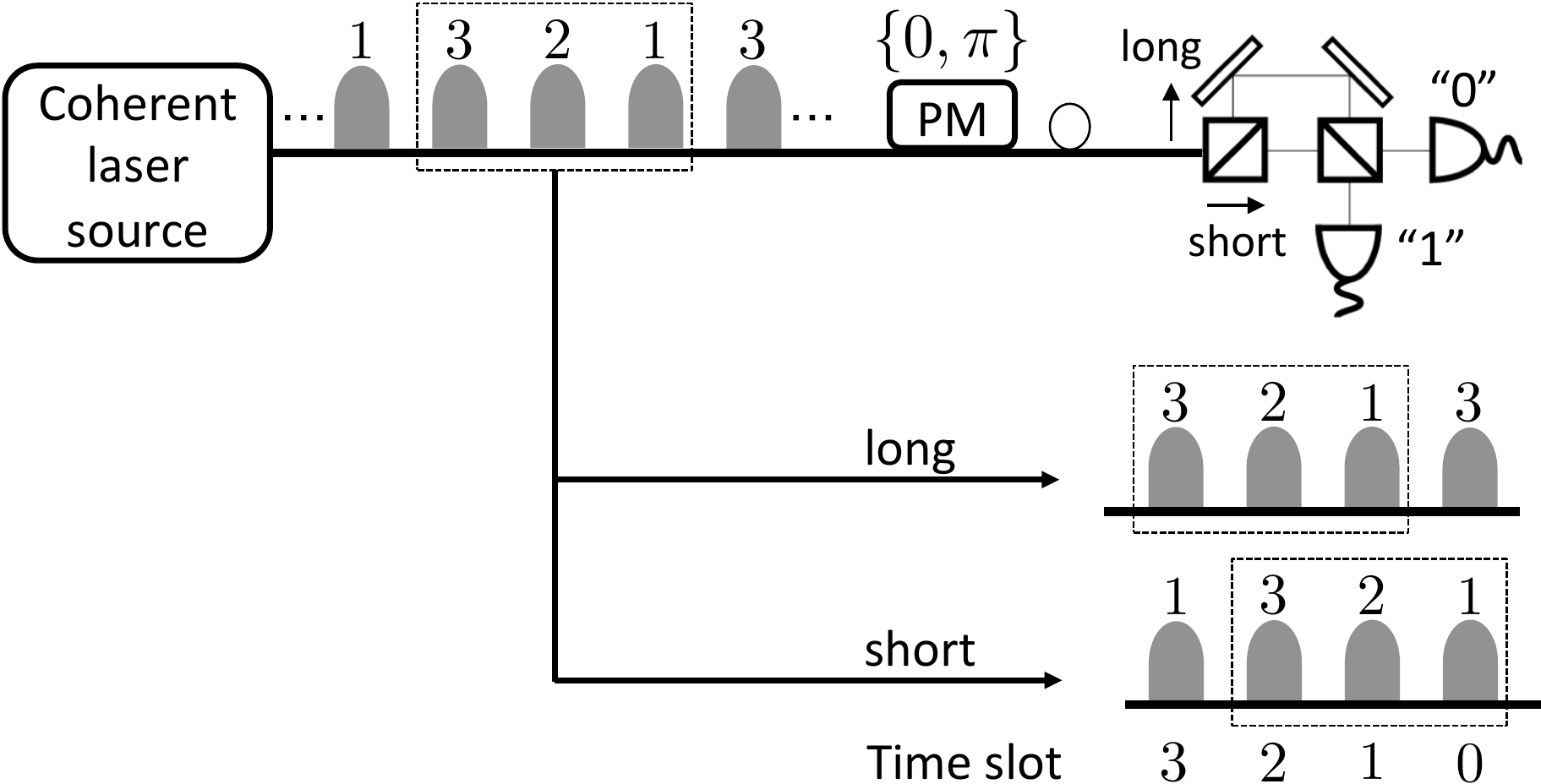}
\caption{
  One possible implementation of the protocol within our security framework. 
  At Alice's site, coherent laser pulse trains are generated by a conventional laser source followed by the phase modulator~(PM)
  that randomly modulates a phase $0$ or $\pi$. 
    At Bob's site, each pulse train is fed to a one-bit delay Mach-Zehnder interferometer with two
    50:50 beam splitters (BSs). 
    The pulse trains leaving the interferometer are measured by two photon-number-resolving (PNR) 
    detectors corresponding to bit values ``0'' and ``1''.
    A successful detection event occurs if Bob detects a single-photon in total among the $1^{\U{st}}$ and $2^{\U{nd}}$ time slots. 
    We emphasize that the use of a coherent laser source and precise control over PM are one of the
    examples of the implementations, and we can use any source as long as it
    satisfies assumptions~(A1)-(A3). 
  }
  \label{fig:actual}
\end{figure}
The protocol runs as follows.
In its description, $|\bm{\kappa}|$ denotes the length of a bit sequence $\bm{\kappa}$.
Fig.~\ref{fig:actual} depicts a protocol with a coherent laser source, which is one possible implementation within our security
framework.
\begin{enumerate}[label=(P\arabic*)]
\setlength{\parskip}{0cm}
\setlength{\itemsep}{0cm}
\item
  Alice chooses a random three-bit sequence $\vec{b}_A$ and sends three pulses in a
  state $\hat{\rho}^{\vec{b}_A}_S$ to Bob via a quantum channel.
  \item
    Bob receives an incoming three pulses and puts them into the Mach-Zehnder interferometer followed by photon detection by using 
    the PNR detectors. We call the event {\it detected} if Bob detects exactly one photon
    in total among the $1^{\U{st}}$ and $2^{\U{nd}}$ time slots. 
    The detection event at the $j^{\th}$ ($1\le j\le 2$) time slot determines the raw key bit $k_B\in\{0,1\}$.
    If Bob does not obtain the detected event, Alice and Bob skip steps (P3) and (P4) below.
  \item
    Bob announces the detected time slot $j$ over an authenticated public channel.
  \item
    Alice calculates her raw key bit $k_A=b^A_j\oplus b^A_{j+1}$.
  \item
    Alice and Bob repeat (P1)-(P4) $N_{\U{em}}$ times. 
  \item
    Alice randomly selects a small portion of her raw key for random sampling. Over the authenticated public channel,
    Alice and Bob compare the bit values for random sampling and obtain the bit error rate $e_{\bit}$ among the sampled bits. 
  This gives the estimate of the bit error rate in the remaining portion. 
\item
  Alice and Bob respectively define their sifted keys $\bm{\kappa}_A$ and $\bm{\kappa}_B$ by concatenating their 
  remaining raw keys. 
  \item
  Bob corrects the bit errors in $\bm{\kappa}_B$ to make it coincide with $\bm{\kappa}_A$ 
  by sacrificing $|\bm{\kappa}_A|f_{\U{EC}}$ bits of encrypted public communication from Alice by
  consuming the same length of a pre-shared secret key.
  \item
    Alice and Bob conduct privacy amplification by shortening their keys by $|\bm{\kappa}_A|f_{\U{PA}}$ to obtain the final keys.
\end{enumerate}
 In this paper, we only consider the secret key rate in the asymptotic limit of an infinite sifted key length. 
 We consider the limit of $N_{\em}\to\infty$ while the following observed parameters are fixed:
\begin{align}
  0\le Q:=\frac{|\bm{\kappa}_A|}{N_{\U{em}}}\le1,~~0\le e_{\U{bit}}\le 1.
  \label{observed}
  \end{align}
\\{\bf Security proof.}
Here, we summarize the security proof of the actual protocol described above and determine the
fraction of privacy amplification $f_{\U{PA}}$ in the asymptotic limit. The proof is detailed in Methods section. 
Our proof is based on the security proof~\cite{Akihiro2017QSTDPS} of the DPS protocol with block-wise phase randomization that employs
complementarity~\cite{Koashi2009}. 
A major difference between our proof and the previous proof~\cite{Akihiro2017QSTDPS}
is that we do not assume block-wise phase randomization.
If block-wise phase randomization is performed, 
the state of each single-block can be seen as a classical mixture of the total photon number state.
This phase randomization simplifies the security proof because the amount of privacy amplification $|\bm{\kappa}_A|f_{\U{PA}}$
can be estimated separately for each photon number emission. 
However, under our assumptions (A1)-(A3), a phase coherence generally exists among blocks,
and the state of each single-block cannot be regarded as a classical mixture of photon number states. 
Therefore, we need to take into account this phase coherence in proving the security. In our security proof, 
the central task is to derive the information increase due to this phase coherence among the blocks. 

For the security proof with complementarity, we consider alternative procedures
for Alice's state preparation in step (P1) and the calculation of her raw key bit $k_A$ in step (P4).
We can employ these alternative procedures to prove the security of the actual protocol because 
Alice's procedure of sending optical pulses, and producing the final key is identical to the actual protocol.
Also, Bob's procedure of receiving the pulses and making his public announcement $j$ (for each round) in the actual 
protocol is identical to the corresponding procedure in the alternative protocol.

As for Alice's state preparation in step~(P1),
she alternatively prepares three auxiliary qubits in system $A_1A_2A_3$,
which remain at Alice's site during the whole protocol, and the three pulses (system $S$) to be sent, in the following state:
\begin{align}
  \ket{\Phi}_{ASR}:=
  2^{-3/2}\bigotimes^3_{i=1}\sum^1_{b^A_i=0}\hat{H}\ket{b^A_i}_{A_i}\ket{\psi_{b^A_i}}_{S_iR_i}.
  \label{coherentLstates}
\end{align}
Here, $\hat{H}:=1/\sqrt{2}\sum_{x,y=0,1}(-1)^{xy}\ket{x}\bra{y}$ is the Hadamard operator, and
$\ket{\psi_{b^A_i}}_{S_iR_i}$ is a purification of $\hat{\rho}^{b^{A}_i}_{S_i}$, namely,
$\tr_{R_i}\ket{\psi_{b^A_i}}\bra{\psi_{b^A_i}}_{S_iR_i}=\hat{\rho}^{b^{A}_i}_{S_i}$.
Note from the assumption (A1) that system $R_i$ is assumed to be possessed by Alice. 

As for the calculation of the raw key bit $k_A$ in step~(P4), this bit can be alternatively extracted by applying
the controlled-not (CNOT) gate on the $j^{\th}$ and
$(j+1)^{\th}$ auxiliary qubits with the $j^{\th}$ one being the control and the $(j+1)^{\th}$ one being the target
followed by measuring the $j^{\th}$ auxiliary qubit in the $X$-basis. 
Here, we define the $Z$-basis states for the $j^{\th}$ auxiliary qubit as $\{\ket{0}_{A_j},\ket{1}_{A_j}\}$, and the CNOT gate
$\hat{U}^{(j)}_{\U{CNOT}}$ is defined on this basis by
$\hat{U}^{(j)}_{\U{CNOT}}\ket{x}_{A_j}\ket{y}_{A_{j+1}}=\ket{x}_{A_j}\ket{x+y~\U{mod}2}_{A_{j+1}}$ with $x,y\in\{0,1\}$.
Also, the $X$-basis states are defined as $\{\ket{+}_{A_j},\ket{-}_{A_j}\}$ with
$\ket{\pm}_{A_j}=(\ket{0}_{A_j}\pm\ket{1}_{A_j})/\sqrt{2}$. 

In order to discuss the security of the key $\bm{\kappa}_A$, we consider a virtual scenario of how well Alice can predict
the outcome of the
measurement complementary to the one to obtain $k_A$. In particular, we take the $Z$-basis measurement as the complementary basis, 
and we need to quantify how well Alice can predict its outcome $z_j\in\{0,1\}$ on the $j^{\th}$ auxiliary qubit. 
  To enhance the accuracy of her estimation, Alice measures the $(j+1)^{\th}$ auxiliary qubit in the $Z$-basis after performing
  $\hat{U}^{(j)}_{\U{CNOT}}$ on the $j^{\th}$ and $(j+1)^{\th}$ auxiliary qubits.
  As for Bob, instead of aiming at learning $\bm{\kappa}_A$, he tries to guess the complementary observable $z_j$ to help Alice's
  prediction. 
  More specifically, Bob performs a virual measurement to learn 
  which of the $j^{\th}$ or $(j+1)^{\th}$ half pulse has a single-photon, whose information is sent to Alice. 
  We define the occurrence of {\it phase error} to be the case where 
  Alice fails her prediction of the complementary measurement outcome $z_j$
  (see Eq.~(\ref{phPOVM}) for the explicit formula of the POVM element of obtaining a phase error). 
Let $N_{\ph}$ denote the number of phase errors, namely, the number of wrong predictions of $z_j$ among $|\bm{\kappa}_A|$ trials. 
Suppose that the upper bound $f(\omega_{\obs})$ on the number of phase errors is estimated as a function of $\omega_{\obs}$ which denotes
all the experimentally available parameters $Q$, $e_{\bit}$ in Eq.~(\ref{observed}) and $\{q_n\}^3_{n=1}$ in Eq.~(\ref{qn}).
  In the asymptotic limit considered here, a sufficient amount of privacy amplification is given by~\cite{Koashi2009}
\begin{align}
Qf_{\U{PA}}=Qh\left(\frac{f(\omega_{\obs})}{|\bm{\kappa}_A|}\right),
\end{align}
where $h(x)$ is defined as $h(x)=-x\log_2x-(1-x)\log_2(1-x)$ for $0\le x\le 0.5$ and $h(x)=1$ for $x>0.5$. 
Then, the secret key rate (per pulse) is given by
\begin{align}
  R=Q\left[1-f_{\U{EC}}-h\left(\frac{f(\omega_{\obs})}{|\bm{\kappa}_A|}\right)\right]/3.
  \label{keyrate}
\end{align}
The quantity $e^{\U{U}}_{\U{ph}}:=f(\omega_{\obs})/|\bm{\kappa}_A|$ in Eq.~(\ref{keyrate}) is the upper bound on the phase error rate
$e_{\U{ph}}:=N_{\ph}/|\bm{\kappa}_A|$.
Our main result, Theorem~\ref{mainth}, derives $e^{\U{U}}_{\ph}$ with experimentally available parameters $Q$, $e_{\bit}$ and
$\{q_n\}^3_{n=1}$ (see Methods section for the proof). 
\begin{theorem}
In  the asymptotic limit of large key length $|\bm{\kappa}_A|$, the upper bound on the phase error rate is given by
\begin{align}
    e^{\U{U}}_{\ph}=\lambda e_{\bit}+\frac{\lambda\sqrt{q_1q_3}+q_2}{Q}
\end{align}
with $\lambda=3+\sqrt{5}$.
\label{mainth}
\end{theorem}
From this theorem and Eq.~(\ref{keyrate}), the scaling of the key rate $R$ with respect to the channel transmission $\eta$ is estimated.
If the protocol is implemented by a weak coherent laser pulse as a light source with its mean photon number $\mu$,
the detection rate $Q$ is in the order of $O(\mu\eta)$ and both $\sqrt{q_1q_3}$ and $q_2$ are in the order of $O(\mu^2)$.
To obtain a positive secret key rate, the upper bound on the phase error rate must be smaller than 0.5: 
\begin{align}
  e^{\U{U}}_{\ph}=\frac{O(\mu^2)}{O(\mu\eta)}=\frac{O(\mu)}{O(\eta)}<0.5.
\end{align}
To maximize the key rate under this constraint, $\mu$ is decreased in proportion to $\eta$. 
Therefore, we find that the scaling of the key rate is in the order of $R=O(\mu\eta)=O(\eta^2)$. 
\\\\{\bf Simulation of secure key rates.}
  \begin{figure}[t]
\includegraphics[width=8.8cm]{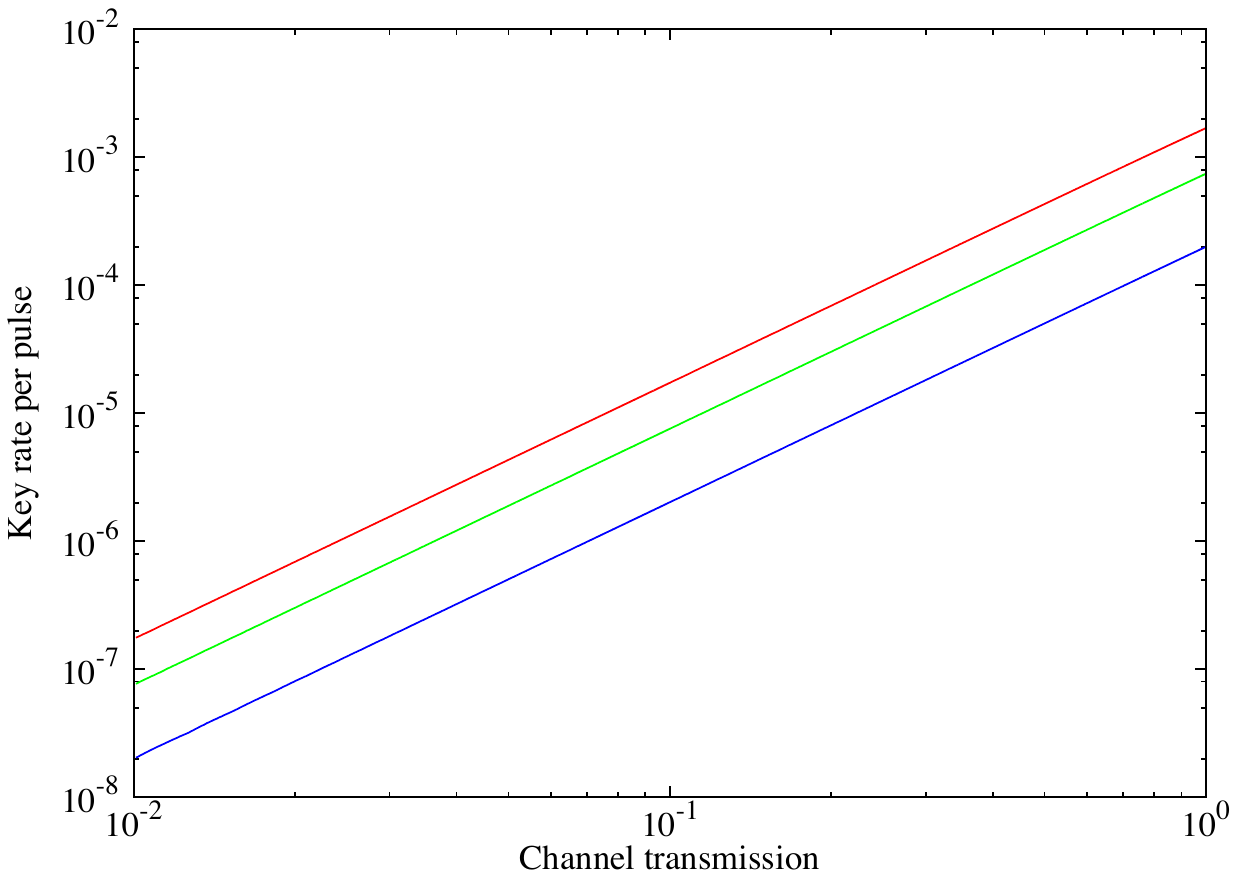}
\caption{Secure key rate $R$ per pulse as a function of the overall channel transmission $\eta$.
  The top, the middle and the bottom lines respectively represent the key rates for $e_{\bit}=0.01$, $e_{\bit}=0.02$ and $e_{\bit}=0.03$. 
}
  \label{fig:keyrate}
  \end{figure}
  We show the simulation results of asymptotic key rate $R$ per pulse given by Eq.~(\ref{keyrate})
  as a function of the overall channel transmission $\eta$  (including detector efficiency). 
  For simplicity of the simulation,
  we assume that each emitted pulse is a coherent pulse from a conventional laser with mean photon number $\mu$.
  In this setting, $q_n$ in Eq.~(\ref{qn}) is given by
\begin{align}
  q_n=\sum^{\infty}_{\nu=n}e^{-3\mu}(3\mu)^{\nu}/\nu!.
  \end{align}
We adopt $f_{\U{EC}}=h(e_{\bit})$ and suppose the detection rate as $Q=2\eta\mu e^{-2\eta\mu}$,
  where in the simulation we omit the cost of random sampling as its cost is negligible in the asymptotic limit. 
In Fig.~\ref{fig:keyrate}, we plot the key rates for $e_{\bit}=0.01$, $e_{\bit}=0.02$ and $e_{\bit}=0.03$ (from top to bottom).
The key rates are optimized over $\mu$ for each value of $\eta$.
The optimized values of $\mu$ when $e_{\bit}=0.02$ is about $7\times 10^{-5}$ (with $\eta=10^{-2}$) and
about $7\times 10^{-3}$ (with $\eta=1$). 
From these lines, we see that all the key rates are proportional to $\eta^2$.
  If we consider the overall channel transmission as $\eta=0.1\times 10^{-0.5\ell/10}$
  (with $\ell$ denoting the distance between Alice and Bob) and laser diodes operating at 1~GHz repetition rate, we can generate 
  a secure key at a rate of 170 bits~s$^{-1}$ for a channel length of
50km and a bit error rate of 1\%.
\\\\{\large \bf Discussion}\\
In this paper, we have provided the information-theoretic security proof of the DPS protocol based on simple source characterizations.
Once one admits the commonly used assumption (A1) to be physically reasonable, 
our proof only requires the independence of the vacuum emission probability from a chosen bit, and the upper bounds $\{q_n\}^3_{n=1}$
on the probabilities that a single-block contains at least $n$ $(n\in\{1,2,3\})$ photons.
Even with these experimentally simple assumptions,
we demonstrated that we can generate a secret key at the rate of about 100 bits~s$^{-1}$ 
for inner-city QKD ($\ell\sim 50$ km) given realistic bit error rate of $1\%\sim 3\%$. 

We end with some open questions. 
As for Alice's side,
it is interesting to extend our security proof by relaxing the assumption (A2) to the case where the vacuum emission probability is
different $\tr\hat{\rho}^{0}_{S_i}\ket{\vac}\bra{\vac}\neq\tr\hat{\rho}^{1}_{S_i}\ket{\vac}\bra{\vac}$, 
and Alice only knows the bounds on $\tr\hat{\rho}^{0}_{S_i}\ket{\vac}\bra{\vac}$ and $\tr\hat{\rho}^{1}_{S_i}\ket{\vac}\bra{\vac}$. 
As for Bob's measurement unit, it is important to relax the assumption (B2) to allow the use of threshold detectors.
\\\\{\large \bf Methods}\\
    {\bf Outline}. Here we prove our main result, Theorem~\ref{mainth}. 
    First, we introduce notations that we use in the following discussions.
    Second, we introduce the POVM (positive operator valued measure) elements corresponding to a bit and phase error. 
    Third, we explain the relation between the $Z$-basis measurement outcome ($z_j$) on the auxiliary qubit system $A_j$ and 
    the number of photons contained in the $j^{\th}$ emitted pulse.
    Finally, we prove Theorem~\ref{mainth} by using two lemmas, Lemmas~\ref{L1P1ephP1} and \ref{lemmacross}. 
    We leave the proofs of Lemmas~\ref{L1P1ephP1} and \ref{lemmacross} to Appendixes A and B, respectively. 
    \\\\
        {\bf Notations.} 
  We first summarize the notations that we use in the following discussions:
  \begin{align}
\hat{P}[\ket{\psi}]:=\ket{\psi}\bra{\psi}
  \end{align}
  for a vector $\ket{\psi}$ that is not necessarily normalized, and the Kronecker delta
    \begin{align}
\delta_{x,y}:=
\begin{cases}
  1 & x=y\\
  0 & x\neq y.
  \end{cases}
    \end{align}
    Furthermore, we introduce the $Z$-basis states of Alice's auxiliary qubit system $A$ as
\begin{align}
  \ket{\bm{z}}_A:=\bigotimes^3_{i=1}\ket{z_i}_{A_i}
  \end{align}
with $\bm{z}=z_1z_2z_3$ and $z_i\in\{0,1\}$, and $\wt(\bm{z})$ denotes the Hamming weight of a bit string $\bm{z}$. 
Let us define the projectors $\hat{P}_a$ (with $0\le a \le 3$), $\hat{P}_{\even}$ and $\hat{P}_{\odd}$ as
\begin{align}
&\hat{P}_a:=\sum_{\bm{z}:\wt(\bm{z})=a}\hat{P}[\ket{\bm{z}}_A],\notag\\
  &\hat{P}_{\even}:=\hat{P}_0+\hat{P}_2,\notag\\
  &\hat{P}_{\odd}:=\hat{P}_1+\hat{P}_3.
\end{align}
    {\bf POVM element for a detected event.}
    We introduce POVM elements for Bob's procedure of determining the detected time slot $j$ and the bit value $k_B$.
    Based on the following procedure, Bob can determine whether the event is detected or not prior to determining
    $j$ and $k_B$. Bob sends the first pulse to the first BS in Fig.~\ref{fig:actual}, and
    after the first pulse is split, one of the pulses goes to the long arm of the Mach-Zehnder interferometer,
    and we call it first half pulse.
      Bob keeps the second pulse as it is, and sends the third pulse to the first BS.
      After the third pulse is split at the first BS,
    one of the pulses goes to the
    short arm of the Mach-Zehnder interferometer (we call it the third half pulse). 
    Bob then performs the quantum nondemolition (QND) measurement of the
    total photon number among the first half pulse, the third half pulse and the second pulse. 
    The detected event is equivalent to an event where the QND measurement reveals exactly one photon. 
    If the detected event occurs, the state of the three pulses after the QND measurement is in the subspace 
    spanned by the orthonormal basis $\{\ket{i}_B\}^3_{i=1}$ with $i$ representing the position of the single-photon
    (at the half pulse when $i=1,3$ and at the original pulse when $i=2$).
    Given the detection, the POVM elements $\{\hat{\Pi}_{j,k_B}\}_{j,k_B}$
  for detecting the bit $k_B$ at the $j^{\th}$ time slot ($1\le j\le 2$)
  is given by
  \begin{align}
    \hat{\Pi}_{j,k_B}=\hat{P}[\ket{\Pi_{j,k_B}}_B]
  \end{align}
  with
  \begin{align}
\ket{\Pi_{j,k_B}}_B&:=\frac{\sqrt{w_j}\ket{j}_B+(-1)^{k_B}\sqrt{w_{j+1}}\ket{j+1}_B}{\sqrt{2}},
\end{align}
where $w_1=w_3=1$ and $w_2=1/2$.
\\\\{\bf Bit- and phase-error POVM elements.} 
Here, we construct the bit- and phase-error POVM elements $\hat{e}_{\bit}$ and $\hat{e}_{\ph}$, respectively.
Alice and Bob measure their systems $A$ and $B$ just after the QND measurement reveals exactly one photon
to learn whether a bit error or phase error exists. 
Importantly, these POVM elements are defined only on Alice's auxiliary qubit system $A$ and Bob's system $B$, and the
assumptions on the encoded states in system $S$ do not come into their description. Therefore, even if the assumptions on Alice's
emitted states are different from those in the security proof~\cite{Akihiro2017QSTDPS} of the DPS protocol with block-wise phase
randomization, the same formulas of the bit- and phase-error POVM elements in~\cite{Akihiro2017QSTDPS} can be used.
Here, we only provide brief explanations of how to construct
$\hat{e}_{\bit}$ and $\hat{e}_{\ph}$, and refer details to Ref.~\cite{Akihiro2017QSTDPS}.

First, we introduce the POVM element $\hat{e}^j_{\U{bit}}$ corresponding to announcing the $j^{\th}$ time slot and the 
occurrence of a bit error.
As a bit error occurs when $k_A$ (Alice's $X$-basis measurement outcome of the $j^{\th}$ auxiliary qubit after
performing $\hat{U}^{(j)}_{\U{CNOT}}$ on the $j^{\th}$ and $(j+1)^{\th}$ ones)
and Bob's measurement outcome $k_B$ are different, $\hat{e}^j_{\U{bit}}$ is given by
  \begin{align}
  &\hat{e}^{j}_{\U{bit}}=(\hat{P}[\ket{++}_{A_jA_{j+1}}]+\hat{P}[\ket{--}_{A_jA_{j+1}}])\otimes\hat{\Pi}_{j,1}\notag\\
  &+(\hat{P}[\ket{+-}_{A_jA_{j+1}}]+\hat{P}[\ket{-+}_{A_jA_{j+1}}])\otimes\hat{\Pi}_{j,0}.
  \label{bitpovm}
  \end{align}
Here and henceforth, we omit identity operators on subsystems, such as those for Alice's  irrelevant auxiliary qubits 
  in the above equation. Eq.~(\ref{bitpovm}) is re-expressed as
  \begin{align}
    &\hat{e}^{j}_{\U{bit}}
    =\sum^1_{s=0}\bigg\{\hat{P}\left[\frac{\ket{00}_{A_jA_{j+1}}+(-1)^s\ket{11}_{A_jA_{j+1}}}{\sqrt{2}}\right]
    \notag\\
    &+\hat{P}\left[\frac{\ket{01}_{A_jA_{j+1}}+(-1)^s\ket{10}_{A_jA_{j+1}}}{\sqrt{2}}\right]\bigg\}
    \otimes\hat{\Pi}_{j,s\oplus1}.
    \label{simplebit}
    \end{align}
  This equation shows that there are no cross terms between even parity terms ($\ket{z_jz_{j+1}}_{A_jA_{j+1}}$ with $z_j+z_{j+1}$ is even) and
  odd parity terms ($\ket{z_jz_{j+1}}_{A_jA_{j+1}}$ with $z_j+z_{j+1}$ is odd). 
  Therefore, we have 
  \begin{align}
    \hat{e}^{j}_{\U{bit}}=\hat{P}_{\even}\hat{e}^{j}_{\U{bit}}\hat{P}_{\even}+\hat{P}_{\odd}\hat{e}^{j}_{\U{bit}}\hat{P}_{\odd}.
    \label{bitoddeven}
  \end{align}
  This equation can be derived more intuitively. 
  The parity of $\wt(\bm{z})$ can be determined by measuring the auxiliary qubits in the $Z$-basis
  except for the $j^{\U{th}}$ one after performing $\hat{U}^{(j)}_{\U{CNOT}}$ on the $j^{\U{th}}$ and $(j+1)^{\U{th}}$
  qubits. 
  This implies that
  the measurement $\{\hat{P}_{\even},\hat{P}_{\odd}\}$ and $\hat{e}^{j}_{\U{bit}}$ commute, and hence we obtain Eq.~(\ref{bitoddeven}).
  Eq.~(\ref{bitoddeven}) plays an important role in proving Theorem~\ref{mainth}. 

  Second, we introduce the POVM element $\hat{e}^j_{\U{ph}}$ corresponding to announcing the $j^{\th}$ time slot and
  the occurrence of a phase error. 
  A phase error event is defined as an event where Alice fails her prediction of the $Z$-basis measurement outcome $z_j$ on the
  $j^{\th}$ auxiliary qubit.
  To enhance the accuracy of her estimation, she measures the $(j+1)^{\U{th}}$ auxiliary qubit
  in the $Z$ basis (with $z_{j+1}$ denoting its result) after performing $\hat{U}^{(j)}_{\U{CNOT}}$, and
  Bob measures system $B$ to learn which of the $j^{\th}$ or $(j+1)^{\th}$ pulse has a single-photon. 
  With the help of information of $z_{j+1}$ and Bob's information, Alice adopts the following strategy for predicting $z_j$. 
  As for the case of $z_{j+1}=1$, if Bob reveals that the $j^{\U{th}}$ [$(j+1)^{\U{th}}$] pulse has a single-photon, Alice predicts
  $z_j=1$ [$z_j=0$].
  On the other hand, if $z_{j+1}=0$, Alice predicts $z_j=0$ regardless of Bob's information.
  The phase error event is defined as an instance of a wrong prediction of $z_j$, and 
  the POVM element corresponding to announcing the $j^{\U{th}}$ time slot and the occurrence of a phase error is represented by
  \begin{align}
    &\hat{e}^j_{\U{ph}}=\sum_{\bm{z}}\hat{P}[\ket{\bm{z}}_A]\notag\\
    \otimes&\left[w_j\delta_{z_{j+1},1}\hat{P}[\ket{j}_B]+
      w_{j+1}\delta_{z_j,1}\hat{P}[\ket{j+1}_B]\right].
    \label{phPOVM}
  \end{align}
  
  Since $\hat{e}^j_{\U{ph}}$ is diagonal in the basis $\ket{\bm{z}}_A$, we have
  \begin{align}
\hat{e}^j_{\U{ph}}=\sum^3_{a=0}\hat{P}_a\hat{e}^j_{\U{ph}}\hat{P}_a.
    \label{ephDiagPa}
    \end{align}
  Then, by taking the sum over all the time slots, we obtain the bit and phase error operators as
  \begin{align}
    \hat{e}_{\U{bit}}:=\sum^{2}_{j=1}\hat{e}^j_{\U{bit}},~\hat{e}_{\U{ph}}:=\sum^{2}_{j=1}\hat{e}^j_{\U{ph}}.
    \label{ebitephOperator}
  \end{align}
    It follows that the probability of having a bit error is given by $\tr\hat{\sigma}\hat{e}_{\bit}$,
    and the one of having a phase error is $\tr\hat{\sigma}\hat{e}_{\ph}$.
    Here, $\hat{\sigma}$ denotes a state of Alice and Bob's
    systems $A$ and $B$ just after the QND measurement reveals
    exactly one photon. 
  \\\\{\bf Relation between $\bm{\wt(z)}$ and $\bm{n_{\bl}}$.}
  Here, we derive the relation between $\wt(\bm{z})$ and $n_{\bl}$.
  For this, we first derive the number of photons ($n_j$) contained in system $S_j$ when $z_j=1$.
  Recall that the assumption (A2) guarantees that
  $\tr\hat{\rho}^{0}_{S_j}\ket{\vac}\bra{\vac}=\tr\hat{\rho}^{1}_{S_j}\ket{\vac}\bra{\vac}=:P^{\vac}_j$.
  From this assumption, by expanding the orthonormal basis of $S_j$ with the photon number states, 
  $\ket{\psi_0}_{S_jR_j}$ (a purification of $\hat{\rho}^{0}_{S_j}$) and
  $\ket{\psi'_1}_{S_jR_j}$ (a purification of $\hat{\rho}^{1}_{S_j}$) can be written as
\begin{align}
 \ket{\psi_0}_{S_jR_j}
  &=\sqrt{P^{\vac}_j}\ket{\vac}_{S_j}\ket{u_0}_{R_j}+\sqrt{P^{1}_{j,0}}\ket{1}_{S_j}\ket{u_1}_{R_j}+\cdots,
    \label{v0}\\
 \ket{\psi'_1}_{S_jR_j}&=\sqrt{P^{\vac}_j}\ket{\vac}_{S_j}\ket{v_0}_{R_j}+\sqrt{P^{1}_{j,1}}\ket{1}_{S_j}\ket{v_1}_{R_j}+\cdots.
  \end{align}
Here, $\ket{u_0}, \ket{u_1}, \ket{v_0}$ and $\ket{v_1}$ are normalized vectors of system $R_j$, and
$P^{1}_{j,b^A_j}:=\tr\hat{\rho}^{b^A_j}_{S_j}\ket{1}\bra{1}$. 
Since a purification has a freedom of choosing a unitary operator $\hat{U}$ on system $R_j$, 
the following state $\ket{\psi_1}_{S_jR_j}$ is also a purification of $\hat{\rho}^{1}_{S_j}$:
\begin{align}
  &\ket{\psi_1}_{S_jR_j}\notag\\
  =&\sqrt{P^{\vac}_j}\ket{\vac}_{S_j}\hat{U}\ket{v_0}_{R_j}+\sqrt{P^{1}_{j,1}}\ket{1}_{S_j}\hat{U}\ket{v_1}_{R_j}+\cdots.
    \label{v1}
\end{align}
In the following discussions, $\hat{U}$ is chosen such that $\hat{U}\ket{v_0}=\ket{u_0}$ holds. 
  Note from Eq.~(\ref{coherentLstates}) that if $z_j=1$, the $j^{\th}$ state can be written as
  $\ket{\Phi_-}_{S_jR_j}:=(\ket{\psi_0}_{S_jR_j}-\ket{\psi_1}_{S_jR_j})/\mathcal{N}$, where $\mathcal{N}$ 
  is an appropriate normalization constant. 
  Using Eqs.~(\ref{v0}) and (\ref{v1}) gives the vacuum emission probability of $\ket{\Phi_-}_{S_jR_j}$ as
  \begin{align}
    \tr\hat{P}[\ket{\Phi_-}_{S_jR_j}]\ket{\vac}\bra{\vac}_{S_j}=0.
    \end{align}
  This equation means that if $z_j=1$, the state in system $S_j$ contains at least one photon. That is,
  \begin{align}
z_j=1\to n_j\ge1.
    \end{align}
  Therefore, we obtain
  \begin{align}
\wt(\bm{z})\ge a\to n_{\bl}=\sum^3_{j=1}n_j\ge a,
    \end{align}
and from Eq.~(\ref{qn}), the following inequality holds
    \begin{align}
      \Pr\{\wt(\bm{z})\ge a\}\le \Pr\{n_{\bl}\ge a\}\le q_a.
      \label{relationZandN}
\end{align}
{\bf Proof of Theorem~\ref{mainth}}.
Here, we prove Theorem 1 in the main text.
For this, 
we first find an upper-bound on the phase error probability $\tr\hat{e}_{\U{ph}}\hat{\sigma}$ in terms of 
the bit error probability $\tr\hat{e}_{\U{bit}}\hat{\sigma}$, which holds for any state $\hat{\sigma}$. 
According to Eq.~(\ref{phPOVM}), since 
$\hat{e}_{\U{ph}}$ is diagonalized in the basis $\ket{\bm{z}}_A$ and
$\hat{P}_0\hat{e}_{\U{ph}}\hat{P}_0=0$, we have
\begin{align}
  \tr\hat{e}_{\U{ph}}\hat{\sigma}
  =&\tr\hat{P}_1\hat{e}_{\U{ph}}\hat{P}_1\hat{\sigma}
  +\sum^3_{a=2}\tr\hat{P}_a\hat{e}_{\U{ph}}\hat{P}_a\hat{\sigma}\notag\\
  \le&\tr\hat{P}_1\hat{e}_{\U{ph}}\hat{P}_1\hat{\sigma}+\sum^3_{a=2}\tr\hat{P}_a\hat{\sigma}.
  \label{fdecom}
\end{align}
To upper-bound $\tr\hat{P}_1\hat{e}_{\U{ph}}\hat{P}_1\hat{\sigma}$ 
with experimentally available data, we employ the following Lemmas~\ref{L1P1ephP1}
and \ref{lemmacross} (see Appendixes A and B for their proofs). 
  \begin{lemma}
    \label{L1P1ephP1}
    \begin{align}
      \hat{P}_1\hat{e}_{\ph}\hat{P}_1\le\lambda\hat{P}_1\hat{e}_{\bit}\hat{P}_1
      \label{eq:l1}
    \end{align}
    with $\lambda:=3+\sqrt{5}$. 
  \end{lemma}
  \begin{lemma}
      \label{lemmacross}
      For any density operator $\hat{\sigma}$,
\begin{align}
    \tr\hat{P}_1\hat{e}_{\bit}\hat{P}_1\hat{\sigma}
    \le\tr\hat{e}_{\bit}\hat{\sigma}+\sqrt{\tr\hat{\sigma}\hat{P}_1\cdot\tr\hat{\sigma}\hat{P}_3}.
          \label{eq:l2}
  \end{align}
  \end{lemma}
  Applying Lemmas~\ref{L1P1ephP1} and \ref{lemmacross} to Eq.~(\ref{fdecom}) leads to 
  \begin{align}
      \tr\hat{e}_{\U{ph}}\hat{\sigma} \le\lambda
      \left(\tr\hat{e}_{\U{bit}}\hat{\sigma}+\sqrt{\tr\hat{\sigma}\hat{P}_1\cdot\tr\hat{\sigma}\hat{P}_3}\right)
      +\sum^3_{a=2}\tr\hat{P}_a\hat{\sigma}.
    \label{eq:afterL1L2}
  \end{align}
  
With the relation between the bit and phase error 
probabilities, the next step is to derive an upper bound on the number of phase errors with experimentally available data. 
For this, we use Azuma's inequality~\cite{Azuma1967} to achieve this goal. 
Suppose that there are $N_{\det}$ detected systems $AB$, and 
Alice and Bob sequentially measure each detected state in order. 
Let us consider the following specific way for choosing the sampled bits among the detected events; 
Alice probabilistically associates each detected event with a sample pair with probability $1-t$ or a code pair with probability $t$,
where $0<t<1$. 
The sample pairs are employed for random sampling to obtain $e_{\bit}$ whereas the code pairs are for 
distilling secret key. 
For each code (sample) pair, Alice and Bob measure their systems to learn whether a phase (bit) error occurs or not.
If a code (sample) pair entails a phase (bit) error, we call such an event ``ph'' (``bit''), otherwise we call
    ``$\overline{{\U{ph}}}$'' (``$\overline{{\U{bit}}}$'').
Also, for each code pair, Alice measures her system $A$ in the $Z$-basis to
obtain the outcome $\wt(\bm{z})=a\in\{0,1,2,3\}$. Such simulataneous measurements are allowed because
$[\hat{e}_{\ph},\hat{P}_a]=0$ holds for any $a$ ($0\le a\le 3$). 
In this stochastic trial, the set of the measurement outcomes for each detected event is given by 
$\mathcal{S}:=\{\bit,\overline{\bit}\}\cup(\bigcup^3_{a=0}\{\ph\wedge a\})\cup(\bigcup^3_{a=0}\{\overline{\ph}\wedge a\})$, and let 
$\xi^i\in\mathcal{S}$ denote the $i^{\U{th}}$ measurement outcome with $1\le i\le N_{\det}$.

Next, let us introduce various parameters that are needed in later discussions. 
The phase error rate in the code pair and the bit error rate in the sample pair are defined as 
\begin{align}
  e_{\ph}=\frac{\sum^3_{a=0}\sum^{N_{\det}}_{i=1}\delta_{\xi^i,\ph\wedge a}}{N_{\U{code}}},~
  e_{\bit}=\frac{\sum^{N_{\det}}_{i=1}\delta_{\xi^i,\bit}}{N_{\sample}},
  \label{maindef}
\end{align}
where $N_{\U{code}}$ and $N_{\sample}$ respectively denote the number of code and sample pairs. 
We define the number $N^l_{\Omega}$ of events that take $\Omega\in \mathcal{S}$ among $l$ trials as
\begin{align}
  N^l_{\Omega}:=\sum^l_{i=1}\delta_{\xi^i,\Omega},
\end{align}
and the sum $P^l_{\Omega}$ of probabilities 
of obtaining $\Omega$ at the $i^{\U{th}}$ trial conditioned on the previous outcomes $\{\xi^k\}^{i-1}_{k=0}$ with $\xi^0$ being constant as
\begin{align}
  P^l_{\Omega}:=\sum^l_{i=1}\Pr\{\xi^i=\Omega|\{\xi^k\}^{i-1}_{k=0}\}.
  \end{align}
We can show that the sequence of random variables
$\{X^0_{\gamma},...,X^{N_{\det}}_{\gamma}\}$ (with $\gamma\in\{\ph,\bit,a\}$), which are defined as 
\begin{align}
    &X^l_{\ph}:=\sum^3_{a=0}(P^l_{\ph\wedge a}-N^l_{\ph\wedge a})\\
    &X^l_{\bit}:=P^l_{\bit}-N^l_{\bit}\\
    &X^l_{a}:=(P^l_{\ph\wedge a}+P^l_{\overline{{\U{ph}}}\wedge a})-(N^l_{\ph\wedge a}+N^l_{\overline{{\U{ph}}}\wedge a})
    \end{align}
and $X^0_{\gamma}:=0$, 
satisfies the Martingale condition with respect to random variables $\{\xi^0,\xi^1,...,\xi^{N_{\det}}\}$, that is 
    $\forall l$, $E[X^l_{\gamma}|\{\xi^k\}^{l-1}_{k=0}]=X^{l-1}_{\gamma}$. 
    Here, $E[X|Y]$ denotes the expectation of $X$ conditioned on $Y$. 
    Also, $\{X^0_{\gamma},...,X^{N_{\det}}_{\gamma}\}$ satisfies a bounded difference condition, namely,
    $\forall l$, $|X^{l}_{\gamma}-X^{l-1}_{\gamma}|\le1$. 
    Once Martingale and the bounded difference conditions are satisfied, we can apply Azuma's inequality; it follows that
    $\forall\zeta>0$ and $\forall N_{\det}>0$
    \begin{align}
      \Pr\{|X^{N_{\U{det}}}_{\gamma}|>N_{\det}\zeta\}\le 2e^{-\frac{N_{\det}\zeta^2}{2}}.
      \label{mainAzuma}
    \end{align}

    Since Eq.~(\ref{eq:afterL1L2}) holds for any $\hat{\sigma}$, by using Cauchy-Schwarz inequality: 
$\sum^m_{i=1}x_iy_i\le\sqrt{\left(\sum^m_{i=1}x_i^2\right)\left(\sum^m_{i=1}y_i^2\right)}$, we have
      \begin{align}
        \frac{\tilde{P}^{N_{\U{det}}}_{\ph}}{t}\le\lambda\left(\frac{P^{N_{\U{det}}}_{\bit}}{1-t}+
        \sqrt{\frac{\tilde{P}^{N_{\U{det}}}_{a\ge 1}}{t}\frac{\tilde{P}^{N_{\U{det}}}_{a=3}}{t}}\right)
        +\frac{\tilde{P}^{N_{\U{det}}}_{a\ge2}}{t},
      \label{neqconazuma}
      \end{align}
      where $\tilde{P}^{N_{\U{det}}}_{\ph}:=\sum^3_{a=0}P^{N_{\U{det}}}_{\ph\wedge a}$ and 
        $\tilde{P}^{N_{\U{det}}}_{a}:=P^{N_{\U{det}}}_{\ph\wedge a}+P^{N_{\U{det}}}_{\overline{{\U{ph}}}\wedge a}$.
    
    By employing the consequence of Azuma's inequality in Eq.~(\ref{mainAzuma}) to each of all the five sums of conditional
    probabilities in Eq.~(\ref{neqconazuma}), we obtain
\begin{align}
  \frac{1}{t}\left(\frac{\tilde{N}^{N_{\U{det}}}_{\ph}}{N_{\U{det}}}-\zeta\right)&
  \le\frac{\lambda}{1-t}\left(\frac{N^{N_{\U{det}}}_{\bit}}{N_{\U{det}}}+\zeta\right)+
  \frac{1}{t}\left(\frac{\tilde{N}^{N_{\U{det}}}_{a\ge2}}{N_{\U{det}}}+\zeta\right)\notag\\
  &+\frac{\lambda}{t}\sqrt{\left(\frac{\tilde{N}^{N_{\U{det}}}_{a=1}}{N_{\U{det}}}
    +\zeta\right)\left(\frac{\tilde{N}^{N_{\U{det}}}_{a=3}}{N_{\U{det}}}+\zeta\right)},
  \label{afterazuma}
\end{align}
where $\tilde{N}^{N_{\U{det}}}_{\ph}:=\sum^3_{a=0}N^{N_{\U{det}}}_{\ph\wedge a}$ and 
        $\tilde{N}^{N_{\U{det}}}_{a}:=N^{N_{\U{det}}}_{\ph\wedge a}+N^{N_{\U{det}}}_{\overline{{\U{ph}}}\wedge a}$.
When $N_{\U{det}}$ gets larger with any fixed $\zeta>0$, 
the probability of violating Eq.~(\ref{afterazuma}) decreases exponentially.
Here and henceforth, we consider the limit of large $N_{\U{det}}$ and neglect $\zeta$. 
In this asymptotic limit, as $N_{\U{code}}\to tN_{\U{det}}$ and $N_{\sample}\to (1-t)N_{\U{det}}$ in Eq.~(\ref{maindef}), we obtain
\begin{align}
e_{\ph}
  \le\lambda e_{\bit}+
  \frac{1}{t}\frac{\tilde{N}^{N_{\U{det}}}_{a\ge2}}{N_{\U{det}}}
  +\frac{\lambda}{t}\sqrt{\frac{\tilde{N}^{N_{\U{det}}}_{a=1}}{N_{\U{det}}}\frac{\tilde{N}^{N_{\U{det}}}_{a=3}}{N_{\U{det}}}}.
  \label{mainazuma}
\end{align}
The last task for deriving the upper bound on $e_{\ph}$ is to upper-bound $\tilde{N}_{a\ge n}^{N_{\U{det}}}$
with experimentally available data.
In so doing, in addition to the detected instances, 
we assume that Alice and Bob randomly associate each of the non-detected instances
with a code instance with probability $t$ or a sample instance with
probability $1-t$. 
Then, we have that the the number $\tilde{N}^{N_{\U{det}}}_{a\ge n}$
of obtaining the outcome $a\ge n$ among the detected instances can never be larger than the one $M^{N_{\U{em}}}_{a\ge n}$
among the {\it emitted} code blocks.
Since the probability 
of obtaining a code pair and the outcome $a\ge n$ when Alice emits the $i^{\U{th}}$ block is 
upper-bounded by $q_n$ according to Eq.~(\ref{relationZandN}),
we can imagine independent trials with probability $q_n$. Therefore, we can use Chernoff bound and obtain
\begin{align}
  \frac{\tilde{N}_{a\ge n}^{N_{\U{det}}}}{N_{\em}}\le\frac{M_{a\ge n}^{N_\em}}{N_{\em}}\le q_{n}+\chi.
  \label{finalineq}
\end{align}
When the number $N_{\em}$ of emitted blocks gets larger for any fixed $\chi>0$, the probability of
violating this inequality decreases exponentially. In the condition of asymptotic limit of $N_{\em}$,
we neglect $\chi$ in the following discussions. 
By substituting Eq.~(\ref{finalineq}) to Eq.~(\ref{mainazuma}), we finally obtain
\begin{align}
    e_{\ph}\le\lambda e_{\bit}+\frac{\lambda\sqrt{q_1q_3}+q_2}{Q}.
\end{align}
This ends the proof of Theorem~\ref{mainth}.\sq
\\\\{\bf  Data availability.}~
No datasets were generated or analyzed during the current study. 
\\{\bf  Acknowledgements.}~
T.S. thanks the support from JSPS KAKENHI Grant Number JP18K13469. 
K.T. thanks the support from JSPS KAKENHI Grant Numbers JP18H05237 
and JST-CREST JPMJCR 1671.
This work was in part supported by Cross-ministerial
Strategic Innovation Promotion Program (SIP) (Council for Science, Technology and Innovation (CSTI)).
\\
  {\bf Competing interests.} The authors declare no competing interests.

\appendix
\section{Proof of Lemma~\ref{L1P1ephP1}}
In this Appendix, we prove Lemma~\ref{L1P1ephP1} in the main text. 
We first explicitly describe $\hat{P}_1\hat{e}_{\bit}\hat{P}_1$ and $\hat{P}_1\hat{e}_{\ph}\hat{P}_1$ 
by respectively using Eqs.~(\ref{simplebit}) and (\ref{phPOVM}) as
\begin{align}
 & \hat{P}_1\hat{e}_{\U{bit}}\hat{P}_1
  =\hat{P}\left[\frac{\ket{001}_{A}}{\sqrt{2}}\right]\otimes\left(\hat{P}[\ket{1}_B]+\frac{1}{2}\hat{P}[\ket{2}_B]\right)\notag\\
  +&\hat{P}\left[\frac{\ket{100}_{A}}{\sqrt{2}}\right]\otimes\left(\frac{1}{2}\hat{P}[\ket{2}_B]+\hat{P}[\ket{3}_B]\right)\notag\\
  +&\sum^1_{s=0}\hat{P}\left[\frac{\ket{010}_{A}+(-1)^s\ket{100}_{A}}{\sqrt{2}}\right]\otimes\hat{\Pi}_{1,s\oplus 1}\notag\\
  +&\sum^1_{s=0}\hat{P}\left[\frac{\ket{001}_{A}+(-1)^s\ket{010}_{A}}{\sqrt{2}}\right]\otimes\Pi_{2,s\oplus 1},
  \label{ebit}
\end{align}
\begin{align}
  \hat{P}_1\hat{e}_{\U{ph}}\hat{P}_1&=(\hat{P}[\ket{001}_{A}]+\hat{P}[\ket{100}_{A}])\otimes\frac{\hat{P}[\ket{2}_B]}{2}\notag\\
  &+\hat{P}[\ket{010}_{A}]\otimes(\hat{P}[\ket{1}]+\hat{P}[\ket{3}_B]).
  \label{expleph}
\end{align}
In Eq.~(\ref{ebit}), it is straightforward to show that
\begin{widetext}
$$
\sum^1_{s=0}\hat{P}\left[\frac{\ket{010}_{A}+(-1)^s\ket{100}_{A}}{\sqrt{2}}\right]\otimes\hat{\Pi}_{1,s\oplus 1}
=\hat{P}\left[\frac{\ket{100}_A\ket{1}_B-\frac{\ket{010}_A\ket{2}_B}{\sqrt{2}}}{\sqrt{2}}\right]+
\hat{P}\left[\frac{\ket{010}_A\ket{1}_B-\frac{\ket{100}_A\ket{2}_B}{\sqrt{2}}}{\sqrt{2}}\right],
$$
$$
\sum^1_{s=0}\hat{P}\left[\frac{\ket{001}_{A}+(-1)^s\ket{010}_{A})}{\sqrt{2}}\right]\otimes\hat{\Pi}_{2,s\oplus 1}
  =\hat{P}\left[\frac{\ket{001}_A\ket{3}_B-\frac{\ket{010}_A\ket{2}_B}{\sqrt{2}}}{\sqrt{2}}\right]+
  \hat{P}\left[\frac{\ket{010}_A\ket{3}_B-\frac{\ket{001}_A\ket{2}_B}{\sqrt{2}}}{\sqrt{2}}\right].
  $$
  \end{widetext}
To upper-bound $\hat{P}_1\hat{e}_{\U{ph}}\hat{P}_1$ by using $\hat{P}_1\hat{e}_{\U{bit}}\hat{P}_1$, 
we remove the four projectors in $\hat{P}_1\hat{e}_{\U{bit}}\hat{P}_1$
that are orthogonal to the range of $\hat{P}_1\hat{e}_{\U{ph}}\hat{P}_1$, which results in
\begin{widetext}
\begin{align}
\hat{P}_1\hat{e}_{\U{bit}}\hat{P}_1
  \ge\frac{1}{2}\left(\hat{P}\left[\frac{\ket{001}_{A}}{\sqrt{2}}\right]+
  \hat{P}\left[\frac{\ket{100}_{A}}{\sqrt{2}}\right]\right)\otimes\hat{P}[\ket{2}_B]
  +
  \hat{P}\left[\frac{\ket{010}_A\ket{1}_B-\frac{\ket{100}_A\ket{2}_B}{\sqrt{2}}}{\sqrt{2}}\right]+
  \hat{P}\left[\frac{\ket{010}_A\ket{3}_B-\frac{\ket{001}_A\ket{2}_B}{\sqrt{2}}}{\sqrt{2}}\right].
  \label{ebit1}
\end{align}
\end{widetext}
Moreover, we apply the following inequality that holds for any normalized vectors $\ket{a}$ and $\ket{b}$ with
$\expect{a|b}=0$
\footnote{Note that 
  Eq.~(\ref{appendixAeq}) holds because the smallest eigenvalue of the following Hermitian operator:
  \begin{align}
    \hat{P}\left[\ket{a}-\frac{\ket{b}}{\sqrt{2}}\right]
    -  \frac{2}{\lambda}\left(\hat{P}[\ket{a}]+\hat{P}\left[\frac{\ket{b}}{\sqrt{2}}\right]\right)
    +\hat{P}\left[\frac{\ket{b}}{\sqrt{2}}\right]
    \label{eqAppendix}
    \end{align}
is zero.
},
\begin{align}
  \hat{P}\left[\ket{a}-\frac{\ket{b}}{\sqrt{2}}\right]\ge
  \frac{2}{\lambda}\left(\hat{P}[\ket{a}]+\hat{P}\left[\frac{\ket{b}}{\sqrt{2}}\right]
    \right)-\hat{P}\left[\frac{\ket{b}}{\sqrt{2}}\right]
\label{appendixAeq}
\end{align}
with $\lambda:=3+\sqrt{5}$, to the last two projectors of the rhs in Eq.~(\ref{ebit1}) and obtain
\begin{align}
  &\hat{P}_1\hat{e}_{\U{bit}}\hat{P}_1
  \ge\frac{1}{2}\left(\hat{P}\left[\frac{\ket{001}_A}{\sqrt{2}}\right]
  +\hat{P}\left[\frac{\ket{100}_A}{\sqrt{2}}\right]\right)\otimes \hat{P}[\ket{2}_B]\notag\\
  +&\frac{1}{\lambda}\left(P[\ket{010}_A\ket{1}_B]+
    P\left[\frac{\ket{100}_A\ket{2}_B}{\sqrt{2}}\right]\right)-\frac{\hat{P}[\ket{100}_A\ket{2}_B]}{4}
\notag\\
    +&\frac{1}{\lambda}\left(P[\ket{010}_A\ket{3}_B]+
    P\left[\frac{\ket{001}_A\ket{2}_B}{\sqrt{2}}\right]\right)-\frac{\hat{P}[\ket{001}_A\ket{2}_B]}{4}
\notag\\
    =&\hat{P}_1\hat{e}_{\U{ph}}\hat{P}_1/\lambda.
  \label{saigo}
\end{align}
This ends the proof of Lemma~\ref{L1P1ephP1}.\sq

\section{Proof of Lemma~\ref{lemmacross}}
In this Appendix, we prove Lemma~\ref{lemmacross} in the main text.
From Eq.~(\ref{bitoddeven}), for any state $\hat{\sigma}$ we have 
\begin{align}
  \tr\hat{P}_{\U{odd}}\hat{e}_{\bit}\hat{P}_{\U{odd}}\hat{\sigma}\le
    \tr\hat{e}_{\bit}\hat{\sigma},
  \label{obs}
\end{align}
which leads to 
\begin{align}
    \tr\hat{P}_1\hat{e}_{\bit}\hat{P}_1\hat{\sigma}\le
  \tr\hat{e}_{\bit}\hat{\sigma}
  -\tr(\hat{P}_1\hat{e}_{\bit}\hat{P}_3+\hat{P}_3\hat{e}_{\bit}\hat{P}_1)\hat{\sigma}.
  \label{main2}
\end{align}
Since 
$-\tr(\hat{P}_1\hat{e}_{\bit}\hat{P}_3+\hat{P}_3\hat{e}_{\bit}\hat{P}_1)\hat{\sigma}\le
|\tr(\hat{P}_1\hat{e}_{\bit}\hat{P}_3\hat{\sigma})|+|\tr(\hat{P}_3\hat{e}_{\bit}\hat{P}_1\hat{\sigma})|
=2|\tr(\hat{P}_1\hat{e}_{\bit}\hat{P}_3\hat{\sigma})|$
\footnote{
The last equality comes from the fact that
    $|\tr\hat{A}|=|\tr\hat{A}^{\dagger}|$ holds for any square matrix $\hat{A}$. 
}, 
we derive an upper bound on $|\tr(\hat{P}_1\hat{e}_{\bit}\hat{P}_3\hat{\sigma})|$. 
From the expression of the POVM element $\hat{e}^j_{\bit}$ given by Eq.~(\ref{simplebit}), we have
\begin{align}
  &\hat{T}:=2\hat{P}_1\hat{e}_{\U{bit}}\hat{P}_3
  =\ket{001}\bra{111}_A\otimes(\hat{\Pi}_{1,1}-\hat{\Pi}_{1,0})\notag\\
  &+\ket{100}\bra{111}_A\otimes(\hat{\Pi}_{2,1}-\hat{\Pi}_{2,0}).
  \label{eq:P1ebitP3}
\end{align}
As $(\hat{\Pi}_{1,1}-\hat{\Pi}_{1,0})^2=(\ket{1}\bra{1}+\ket{2}\bra{2})/2$ and
$(\hat{\Pi}_{2,1}-\hat{\Pi}_{2,0})^2=(\ket{2}\bra{2}+\ket{3}\bra{3})/2$, we obtain
\begin{align}
\hat{T}^{\dagger}\hat{T}=&\hat{P}[\ket{111}_A]\otimes[(\hat{\Pi}_{1,1}-\hat{\Pi}_{1,0})^2+(\hat{\Pi}_{2,1}-\hat{\Pi}_{2,0})^2]\notag\\
\le&\hat{I}_{AB}.
\end{align}
This inequality implies that the operator norm of $\hat{T}$ is upper-bounded by 1:
\begin{align}
  ||\hat{T}||_{\infty}:=\min\{c\ge 0~\U{s.t.}~\forall v~||\hat{T}v||\le c||v||\}\le1,
  \label{operatorineq}
\end{align}
where $||\cdot||:=\sqrt{\expect{\cdot|\cdot}}$. 
Next, we define
\begin{align}
\hat{G}:=\hat{P}_3\hat{\sigma}\hat{P}_1.
\end{align}
Its trace norm $||\hat{G}||_1$ is written by using a unitary operator $\hat{W}$ and is calculated as
\begin{align}
  ||\hat{G}||_1&=|\tr\hat{G}\hat{W}|
  =|(\sqrt{\hat{\sigma}}\hat{P}_3,\sqrt{\hat{\sigma}}\hat{P}_1\hat{W})|\notag\\
  &\le\sqrt{(\sqrt{\hat{\sigma}}\hat{P}_3,\sqrt{\hat{\sigma}}\hat{P}_3)}
  \sqrt{(\sqrt{\hat{\sigma}}\hat{P}_1\hat{W},\sqrt{\hat{\sigma}}\hat{P}_1\hat{W})}\notag\\
  &=\sqrt{\tr\hat{P}_3\hat{\sigma}}\sqrt{\tr\hat{P}_1\hat{\sigma}},
  \label{traceineq}
\end{align}
where we use the definition of Hilbert-Schmidt inner product in the second equality and use Schwarz inequality in the first inequality. 
Finally, using H\"older's inequality, Eqs.~(\ref{operatorineq}) and (\ref{traceineq}) gives
\begin{align}
  2|\tr(\hat{P}_1\hat{e}_{\bit}\hat{P}_3\hat{\sigma})|
  &=|\tr\hat{T}\hat{G}|
  \le||\hat{T}\hat{G}||_1
  \le||\hat{T}||_{\infty}||\hat{G}||_1\notag\\
  &\le\sqrt{\tr\hat{P}_3\hat{\sigma}\cdot\tr\hat{P}_1\hat{\sigma}}.
      \label{final1}
\end{align}
Therefore, we obtain
\begin{align}
  -\tr(\hat{P}_1\hat{e}_{\bit}\hat{P}_3+\tr\hat{P}_3\hat{e}_{\bit}\hat{P}_1)\hat{\sigma}
  \le\sqrt{\tr\hat{P}_1\hat{\sigma}\cdot\tr\hat{P}_3\hat{\sigma}}.
  \label{final}
\end{align}
Finally, by using Eqs.~(\ref{main2}) and (\ref{final}), we conclude that 
\begin{align}
  \tr\hat{P}_1\hat{e}_{\bit}\hat{P}_1\hat{\sigma}
  \le\tr\hat{e}_{\bit}\hat{\sigma}+\sqrt{\tr\hat{\sigma}\hat{P}_1\cdot\tr\hat{\sigma}\hat{P}_3}.
  \end{align}
This ends the proof of Lemma~\ref{lemmacross}. \sq


\end{document}